\documentclass[doublecol,amsmath,amssymb]{epl2}

\usepackage{graphicx}
\usepackage{dcolumn}
\usepackage[dvipsnames]{xcolor}

\begin{document}

\title{Non-Thermal Einstein Relations}

\author{Robin Guichardaz$^1$, Alain Pumir$^1$  and Michael Wilkinson$^2$,}

\institute{$^1$ Univ Lyon, Ens de Lyon, Univ Claude Bernard, CNRS, Laboratoire de Physique, F-69342 Lyon, France\\ 
 $^2$ Department of Mathematics and Statistics,
The Open University, Walton Hall, Milton Keynes, MK7 6AA, England
}

\pacs{05.10.Gg}{Stochastic analysis methods (Fokker-Planck, Langevin, etc.)}
\pacs{05.40.-a}{Fluctuation phenomena, random processes, noise, and Brownian motion}

\abstract{
We consider a particle moving with equation of motion $\dot x=f(t)$, 
where $f(t)$ is a random function with statistics which are independent 
of $x$ and $t$, with a finite drift velocity $v=\langle f\rangle$ and in the presence
of a reflecting wall. Far away from the wall, translational invariance implies that the stationary 
probability distribution is $P(x)\sim \exp(\alpha x)$.  A  classical 
example of a problem of this type is sedimentation equilibrium, where
$\alpha$ is determined by temperature. In this work 
we do not introduce a thermal reservoir and $\alpha$ is determined 
from the equation of motion. We consider a general approach 
to determining $\alpha$ which is not always in agreement with Einstein's 
relation between the mean velocity and the diffusion 
coefficient. We illustrate our results with a model inspired by
the Boltzmann equation.
}

\maketitle

\section{Introduction}
\label{sec: 1}

This Letter discusses a new perspective on a classic problem of 
statistical physics. Consider the motion of a particle with equation 
of motion
\begin{equation}
\label{eq: 1}
\frac{{\rm d}x}{{\rm d}t}=f(t)
\end{equation}  
where $f(t)$ is a random function, with statistical properties
which are independent of $x$ and of $t$. 
We might wish to characterise the probability distribution of the coordinate 
$x(t)$. 
If we seek a probability distribution which is stationary in time, 
this distribution should respect, away from the boundaries, 
the translational invariance of the problem.
The stationary probability density must have an exponential form
\begin{equation}
\label{eq: 3}
P(x)=A\,\exp(\alpha x)
\end{equation}
where $A$ is a normalisation constant. 
In this Letter we present a general 
formula determining the exponent $\alpha$ in terms of the statistics of the 
function $f(t)$. 

This problem is closely related to the classical treatment of sedimentation 
equilibrium~\cite{Perrin:1913} by Einstein \cite{Einstein:05} and Sutherland \cite{Sutherland:05}, 
who used statistical mechanics 
to describe the particle motion, in terms of a diffusion process,
and to relate $\alpha $ to temperature via the diffusion coefficient.
In this work we treat equation (\ref{eq: 1}) as a purely dynamical 
process, and the exponential solution (\ref{eq: 3}) is a consequence of 
translation symmetry, rather than thermal equilibrium. We are concerned with 
the relation
between $\alpha$ and dynamical quantities.  In a homogeneous system,
we expect the motion at long time to resemble a biased random walk 
with drift velocity $v$ and diffusion coefficient $D$, given by
\begin{equation}
\label{eq: 2}
v=\langle f(t)\rangle
\ ,\ \ \ 
D=\frac{1}{2}\int_{-\infty}^\infty{\rm d}t\ \langle (f(t) - v ) (f(0) - v)\rangle
\end{equation}
where $\langle X\rangle$ denotes the expectation value of $X$ throughout. 
We assume that both $v$ and $D$ are finite, and non zero.
In the case of a Brownian particle in a thermal bath, an appropriate description of the evolution of the 
probability distribution function (PDF), $P(x,t)$, is given by the Fokker-Planck
equation:
\begin{equation}
\label{eq:FP}
\frac{\partial P}{\partial t}=-\frac{\partial}{\partial x}\left(vP\right)+D\frac{\partial^2 P}{\partial x^2} \ .
\end{equation}
Seeking a stationary solution of Eq.~(\ref{eq:FP}) with an exponential dependence on
$x$,  Eq.~(\ref{eq: 3}), of Eq.~(\ref{eq:FP}) leads to an explicit formula
for $\alpha$: 
\begin{equation}
\label{eq: 5}
\alpha_{\rm E}=\frac{v}{D}
\ .
\end{equation}
In the case of sedimentation equilibrium, the exponent of the exponential
distribution is determined by the temperature, and equation (\ref{eq: 5}) is 
the basis of the relation between mobility, diffusion coefficient and 
temperature which was introduced by Einstein \cite{Einstein:05}
and Sutherland \cite{Sutherland:05}.
In the remainder of this text we refer to (\ref{eq: 5}) as the classical 
Einstein relation, although we do not consider a coupling with a thermal bath.

In general, the evolution of the probability density for the system 
is not always faithfully represented by (\ref{eq:FP}).
Although deriving the proper formulation is a challenging
task (a variety of different approaches are 
discussed in \cite{vanK+61,Moy+49,Fox+78,Wio+89,San+86}),
our approach in this letter does {\it not} make explicit use of
a generalisation of Eq.~(\ref{eq:FP}), but rather uses large deviation 
theory~\cite{Fre+84,Touchette:09}. 
As a consequence of the fact that Eq.~(\ref{eq:FP}) is no more than an 
approximation, the status of Eq.~(\ref{eq: 5}) is uncertain. 

We analyse a simple model, 
where $f(t)$ is \emph{telegraph noise}, and we determine a closed
form for $\alpha$, which differs from Eq.~(\ref{eq: 5}).
In the telegraph noise model, a particle moves with one of two
possible velocities, and the transition between the two velocity states 
is completely random.  

Because the exponent $\alpha$ is a very fundamental characterisation of 
the simple dynamical process (\ref{eq: 1}), we provide a general analysis
of this quantity. We show that large deviation theory provides a powerful approach to deriving a 
\emph{generalised Einstein relation}, in the form of an implicit equation for
$\alpha$ in terms of cumulants of $f(t)$. Equation (\ref{eq: 5}) appears as an 
approximation of this general expression in the case where the 
random process $f$ is described by a Gaussian process. 
The application of the 
general formula derived from large deviation theory is illustrated 
here by using the telegraph noise model as an example.

Finally, we discuss how deviations from Eq.~(\ref{eq: 5}) 
could affect the sedimentation equilibrium. 
The telegraph noise process can be viewed as a simplified model for 
the microscopic motion of molecules in gases, in which there are only 
two possible velocities. The analysis is readily extended to the Boltzmann 
equation, where atoms move ballistically between collisions, which occur at 
random intervals and result in an instantaneous change in the velocity of a 
particle. In the general case the exponent $\alpha$ is not given 
correctly by (\ref{eq: 5}) for a sedimentation equilibrium described 
by the Boltzmann equation. This raises a question about the validity of the 
classical Einstein relation for sedimentation equilibrium, and potentially
for other physical processes. 
In the limit where the suspended particles are very massive 
compared to the gas molecules, however, we notice that the collision  
term in the Boltzmann equation is replaced by a diffusion term in the particle 
velocity. We show that for this model equation (\ref{eq: 5}) is exact, so 
that the classical Einstein 
relation is valid for the sedimentation equilibrium of macroscopic particles. 

\section{Telegraph noise model}
\label{sec: 2}

We first discuss the example where the velocity $f(t)$ in 
Eq.~(\ref{eq: 1}) is a random telegraph noise. Namely, we assume that $f(t)$
can be either of the two values $f_+$ and $f_-$. The system switches 
from $f_+$ to $f_-$ (respectively $f_-$ to $f_+$) with transition rates
$R_+$ (respectively $R_-$). 
The probability in the steady state regime of the velocity to be $f_+$ ($f_-$),
$p_+$ (respectively $p_-$), is simply
given by $p_+ = R_-/(R_- + R_+)$ ($p_- = R_+/(R_- + R_+)$). As
a consequence, the mean velocity $\langle f \rangle$ is given by
\begin{equation}
\langle f \rangle = \frac{ R_+ f_- + R_- f_+}{ R_+ + R_-}
\ . 
\end{equation}
We assume the presence of an impervious wall, say at $x = 0$, and require
that the
two velocities $f_+$ and $f_-$ to be of opposite signs, which is required
to impose zero flux boundary condition at the wall.
In fact, in order to reach a stationary state, the zero flux condition is needed everywhere. 
Without any loss of generality, we assume that $f_+ > 0$, $f_- < 0$, and that
the averaged velocity $\langle f \rangle $ is negative. 

We introduce the probability $P_{+}(x,t)$ 
($P_-(x,t)$) that 
the position is $x$ at time $t$, the velocity of the system being $f_+$
($f_-$). The evolution equation for $P_+$, $P_-$ is simply:
\begin{equation}
\label{eq:evol_P}
\frac{\partial}{\partial t} 
\begin{pmatrix}{P_+ \cr P_-}\end{pmatrix}
= - \frac{\partial}{\partial x} 
\begin{pmatrix}{f_+ P_+ \cr f_- P_-}\end{pmatrix}+ 
\begin{pmatrix}{ - R_+ & R_- \cr R_+ & -R_- }\end{pmatrix}
\begin{pmatrix}{P_+ \cr P_-}\end{pmatrix}
\ .
\end{equation}
Steady-state solutions of the form 
$P_{+,-}(x) \propto A_{+,-} \exp( \alpha x )$, consistent with
Eq.~(\ref{eq: 3}), can be readily found by imposing that the matrix $M(-\alpha)$,
defined by
\begin{equation}
M(-\alpha ) \equiv
\left(\begin{array}{cc}
-\alpha f_+ - R_+ & R_- \cr
R_+  & - \alpha f_- - R_-
\end{array}\right)
\label{eq:def_M}
\end{equation}
has a zero determinant: $\det ( M( \alpha ) ) = 0$.
This condition leads to a simple algebraic equation, with only 
one non-zero
root: 
\begin{equation}
\alpha = - \frac{ f_+ R_- + f_- R_+ }{f_+ f_- }
\ .
\label{eq:alpha_sed}
\end{equation}
With our assumptions for the signs, the exponent $\alpha $ is negative. 
More generally, the product $\langle f \rangle \times \alpha > 0$.
This guarantees that away from the reflecting wall, 
the solution decays exponentially, similar to what happens in the 
sedimentation problem of Brownian particles~\cite{Einstein:05,Sutherland:05}. 

The value of $\alpha$ given by Eq.~(\ref{eq:alpha_sed}), however, differs
from the prediction given by Eq.~(\ref{eq: 5}). From the solution of Eq.~(\ref{eq:evol_P}), 
in the homogeneous case ($\partial/\partial_x \rightarrow 0$), one determines that the correlation function decays
exponentially with rate $R_- +R_+$, and by computing the variance of $f$ we obtain
the correlation function
\begin{eqnarray} 
&&\langle (f(t) - \langle f \rangle ) (f(0) - \langle f \rangle ) \rangle 
\nonumber \\
&=&\frac{R_+ R_-(f_+ - f_-)^2}{(R_+ + R_-)^2} \exp[ - (R_+ + R_-) t] 
\end{eqnarray}
so the diffusion coefficient $D$ is equal to 
\begin{equation}
D = \frac{ R_+ R_-}{(R_+ + R_-)^3} (f_+ - f_-)^2
\ .
\end{equation}
The resulting ratio 
$v/D$ clearly differs from the expression for $\alpha$, 
Eq.~(\ref{eq:alpha_sed}), thus calling for a revisiting of the 
Einstein-Sutherland relations. We find
\begin{equation}
\frac{\alpha}{\alpha_{\rm E}}=-\frac{R_- R_+}{(R_-+R_+)^2}
\frac{(f_+-f_-)^2}{f_+f_-} 
\ .
\end{equation}
In general, this ratio may be either very large or very small. 
After some algebra, it can be shown that the ratio approaches unity
whenever the dimensionless parameter
\begin{equation}
	\mu = \frac{\langle f \rangle}{D}\frac{f_+-f_-}{R_++R_-}
\label{eq:mu}
\end{equation}
becomes very small. 
The quantity $\mu$ can be rewritten as $\mu = \alpha_{\rm E} \ell $,
where $\alpha_{\rm E}$ is given by Eq.~(\ref{eq: 5}), and $\ell$ is
effectively the mean free path of the particle. The length $\ell$ is the 
product of $1/(R_+ + R_-)$, which provides an estimate of how long the particle
stays with either velocity $f_+$ or $f_-$, and of $f_+ - f_- = 
(f_+ - \langle f \rangle ) - (f_- - \langle f \rangle )$, which
is the size of the difference between the mean and the instantaneous
velocity. Thus, $\ell$ is of the order of the size travelled by a particle 
between two collisions, hence the mean free path interpretation.
Thus, the condition $\mu \rightarrow 0$ expresses that the mean free path,
$\ell$, is much smaller than the typical decay length predicted by Einstein
theory.

Note that the solution $\alpha=0$ is formally always valid. 
It corresponds to the 
homogeneous case, where the density of probability is uniform, and thus,
non-normalisable.

\section{A general form for the Einstein relation}
\label{sec: 3}

To proceed, we now consider the general problem described
by Eq.~(\ref{eq: 1}). We consider the integral of equation (\ref{eq: 1})
\begin{equation}
\label{eq: 3.1}
x(t)=x(0)+\Delta x(t)
\ , \ \ \ \Delta x(t)=\int_0^t \mathrm{d}t'\ f(t') 
\ .
\end{equation}
Let $\pi(\Delta x,t)$ be the probability density of $\Delta x$ at time $t$.
We express the condition that the distribution $P(x)$ is stationary in the form:
\begin{equation}
P(x ) = \int \mathrm{d} \Delta x\ P(x - \Delta x) \pi (\Delta x,t)
\ .
\label{eq:stat_pdf}
\end{equation}
Using explicitly the exponential form of the PDF $P(x)$, Eq.~(\ref{eq: 3}),
one obtains the expression:
\begin{equation}
\int \mathrm{d} \Delta x\ \exp( - \alpha \Delta x) \pi(\Delta x) = 1
\ .
\label{eq:large_dev_0}
\end{equation}
Eq.~(\ref{eq:large_dev_0}) can be interpreted as the average of 
$\exp(- \alpha \Delta x)$, the variable $\Delta x(t)$ being characterized
by its PDF, $\pi(\Delta x,t)$. 
It is valid provided $t$ is much larger than the correlation
time of the original process $f(t)$ so that we can assume that
$\Delta x(t)$ is independent of $x$. This gives 
\begin{equation}
\left\langle \exp \left( - \alpha \int_0^t \mathrm{d}t'\ f(t') \right)  \right\rangle = 1 
\ .
\label{eq:large_dev}
\end{equation}
In the $t \rightarrow \infty$ limit, the large deviation principle~\cite{Touchette:09} provides 
an appropriate approach.
We introduce here the scaled cumulant generating function~\cite{Touchette:09},
$\lambda(k)$, defined by:
\begin{equation}
\lambda(k) \equiv \lim_{T \rightarrow \infty} \frac{1}{T} \ln \Big\langle \exp \Big(k\int_0^T \mathrm{d}t'\ f(t')  \Big)  \Big\rangle 
\label{eq:def_lambda}
\end{equation}
which describes the exponential growth of the average 
$\Big \langle \exp( k \int_0^T \mathrm{d}t'\ f(t')) \Big \rangle $ as a function
of time $T$. The condition Eq.~(\ref{eq:large_dev}) merely states that 
\begin{equation}
\label{eq: 3.2}
\lambda(-\alpha) = 0
\ .
\end{equation}
Thus, the determination of spatial distribution of
particles in a sedimentation equilibrium amounts to finding solutions 
of Eq.~(\ref{eq: 3.2}),
which is a simple condition for $\alpha$ that can be simply applied if the 
cumulant generating function can be determined.

\begin{figure}[h!]
\vspace{-0.5cm}
\includegraphics[width=0.45\textwidth]{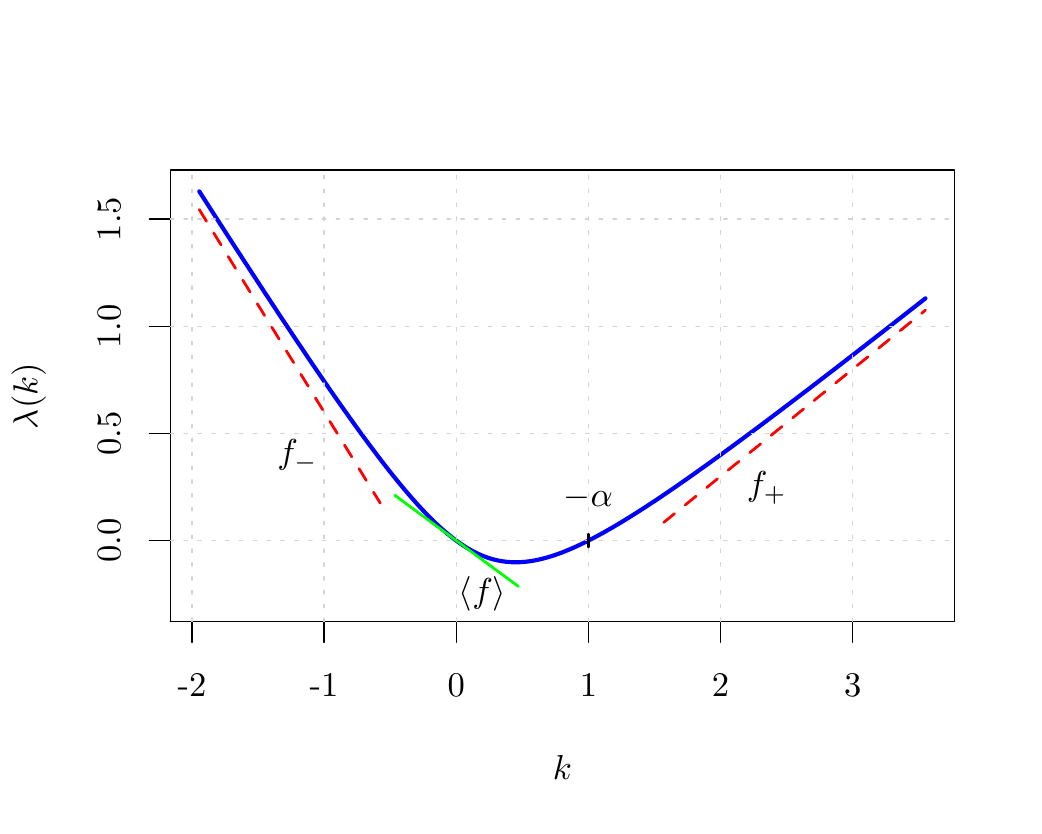}
\vspace{-0.5cm}
\caption{
(Colour online). In blue, plot of $\lambda(k)$.
The parameters are $R_-=0.4$, $R_+=0.7$, $f_-=-1$, and $f_+=0.5$, 
leading to $\langle f \rangle\simeq-0.45<0$. 
We remark that $\lambda'(0)=\langle f \rangle$ (green line),
and that the slope of the asymptote in $k\to-\infty$ ($k\to+\infty$) is $f_-$ ($f_+$) (red dotted lines). 
Moreover, one has $\alpha\langle f\rangle>0$.
}
\label{fig:lambda_k}
\end{figure}

We now illustrate the application of the large deviation theory approach by 
using (\ref{eq: 3.2}) to determine $\alpha$ for the telegraph 
noise model. To this end, we discretize time, and consider $f_n = f(n \Delta t)$ and 
$x_n = x(n \Delta t)$, where $\Delta t$ is a very small time step. Following 
the large deviation approach, we consider the function $\lambda(k)$, defined by
Eq.~(\ref{eq:def_lambda}). To evaluate $\lambda(k)$, we adapt the 
general approach described in~\cite{Touchette:09} (see in particular Section 4.3) 
as follows. With the telegraph noise process,
$f_n$ can take only two values, $f_+$ and $f_-$, so the integral
in Eq.~(\ref{eq:def_lambda}) reduces (up to an overall factor $\Delta t$)
to a sum of terms equal to $f_+$ and $f_-$, depending on the 
state of the system. The expectation value is computed by summing 
over all sequences $f_1,f_2,\ldots,f_n,\ldots$. Because the steps are 
statistically independent, the probability density for a sequences of steps
may be expressed as a product of the form $\prod_j P(f_{j+1},f_j)$, where
$P(f_{j+1},f_j)$ is the probability to reach $f_{j+1}$ at $t_j+\Delta t$, if 
the particle is in velocity state $f_j$ at time $t_j$. The summation over all
possible values of $f_j$ can be represented as a product of a string of 
matrices (which are $2\times 2$ matrices, because the telegraph noise
model has only two possible velocities at each time step).
The quantity $\langle \exp( k \Delta t \sum_{i=0}^n f_i ) \rangle$
grows exponentially as a function of $n$ as $\xi(k)^n$, where
$\xi(k)$ is the largest eigenvalue of the \lq tilted' transition 
matrix~\cite{Touchette:09}, given by:
\begin{equation}
\Pi_k = \left(\begin{array}{cc} 
(1 - R_+ \Delta t)~ \mathrm{e}^{k f_+ \Delta t} & R_- \Delta t~ \mathrm{e}^{k f_+ \Delta t} \cr
R_+\Delta t ~\mathrm{e}^{k f_- \Delta t} & (1 - R_- \Delta t) ~\mathrm{e}^{k f_- \Delta t} 
\end{array}\right)
\ .
\label{eq:tilt_transition_mat}
\end{equation}
Thus, $\lambda(k) $ reduces to the logarithm of the largest value of
$\Pi_k$. In the limit $\Delta t \rightarrow 0$, the matrix $\Pi_k$
reduces to a sum of the identity matrix, $\rm{Id}$, plus $\Delta t $ times
the matrix $M(k)$, defined by Eq.~(\ref{eq:def_M}).
From this simple representation of the matrix $\Pi_k$, it immediately 
follows that the values of $\alpha$ for which 
$\xi(-\alpha) = 1$ in the limit $\Delta t \rightarrow 0$ are exactly the values 
of $\alpha$ for which $\det( M(-\alpha) ) = 0$, thus establishing that
$\alpha$ can be in fact established using large deviation theory.
The function $\lambda(k)$ for the telegraph noise model is illustrated in 
Fig. \ref{fig:lambda_k}.

Equation (\ref{eq: 3.2}) provides a simple criterion to determine $\alpha$ if 
the cumulant generating function $\lambda(k)$ can be determined. In many cases, this 
will not be practicable, and it is desirable to have an alternative approach.
To proceed further, we notice that the expression Eq.~(\ref{eq:large_dev})
can be simply written as a series in powers of $\alpha$, in the form:
\begin{eqnarray}
\lambda(-\alpha) & = & \lim_{T\rightarrow \infty} \frac{1}{T} \ln
\Big \langle \sum_{n=0}^\infty \frac{ (-\alpha)^n}{n!} \Bigl( \int_0^T \mathrm{d}t\ f(t) \Bigr)^n \Big \rangle \nonumber \\
& = & \sum_{n=0}^\infty \frac{(-1)^n}{n!} c_n \alpha^n 
\label{eq:cum_exp}
\end{eqnarray}
where $c_n$ are defined as the integrals of the $n^{\rm th}$ order
cumulants of the distribution of $f(t)$:
\begin{equation}
c_n = \lim_{T \rightarrow \infty} \frac{1}{T} \int_0^T \mathrm{d}t_1 \cdots \int_0^T \mathrm{d}t_n
\,\kappa[ f(t_1), \ldots, f(t_n) ]
\ .
\label{eq:def_cn}
\end{equation}
The first cumulants are simply 
\begin{eqnarray} 
& \kappa & \!\!\![ f(t_1) ] = \langle f(t_1) \rangle \nonumber \\
& \kappa & \!\!\![ f(t_1), f(t_2) ] = \langle f(t_1) f(t_2) \rangle - \langle f(t_1) \rangle^2 \nonumber \\
& \kappa & \!\!\![ f(t_1), f(t_2), f(t_3) ] = \langle f(t_1) f(t_2) f(t_3) \rangle \nonumber \\
& & ~~~ - \langle f(t_1) \rangle \langle f(t_2) f(t_3) \rangle  
- \langle f(t_2) \rangle \langle f(t_1) f(t_3) \rangle \nonumber \\
& &  ~~~ - \langle f(t_3) \rangle \langle f(t_1) f(t_2) \rangle + 2 \langle f \rangle^3 
\ .
\label{eq:def_cumul}
\end{eqnarray}
It is straightforward to check that the coefficients $c_1$ and $c_2$, as 
defined by Eq.~(\ref{eq:def_cn}) coincide with $\langle f \rangle$
and $D$, as defined by Eq.~(\ref{eq: 2}). This immediately shows that the
Einstein-Sutherland relations are exact when the cumulants of order
higher than $3$ vanish, which is the case when $f$ is given by
a Gaussian process. This conclusion does not depend on whether the 
process is Markovian or not. 

Finally, in the telegraph model case, the parameter $\mu$, 
defined in Eq.~(\ref{eq:mu}), effectively specifies how far the process 
is from being Gaussian.
Specifically, the deviation from a Gaussian distribution in 
Eq.~(\ref{eq:cum_exp}) are due to the terms $c_n$ for $n > 2$.
One therefore has to compare the relative importance of $c_n\alpha_{\rm E}^n/n!$ for $n>2$ 
with $c_1\alpha_{\rm E}$ 
(or, equivalently, with $c_2\alpha_{\rm E}^2/2$).
One can show that for $n>2$
\begin{equation}
	\frac{c_n\alpha_{\rm E}^n}{n!\, c_1 \alpha_{\rm E}} = \mu^{n-2}G_n\left(\frac{R_-}{R_+}\right)
\end{equation}
where the $G_n$ are bounded functions, which implies that the solution 
of $\lambda(\alpha) = 0$ in the limit $\mu\to0$
tends to $\alpha = \alpha_{\rm E}$, thus justifying the Einstein equation.

\section{More refined models of sedimentation}
\label{sec: 4}

Our observation that the exponent for sedimentation equilibrium in
the case of a telegraph noise model does not agree
with the classical Einstein relation 
raises the question as to whether the discrepancy exists in more refined 
models. 

The telegraph noise model is close in structure to the Boltzmann 
model for the motion of atoms in a dilute gas, where the atoms move ballistically between 
collisions, and have their velocities changed discontinuously at collision events which 
occur at random times. The difference is that the Boltzmann equation has a continuum
of allowed velocities, so that the probability density is a function of a continuous velocity 
$v$ and the probability density $P(x,v,t)$ satisfies a version of the Boltzmann equation 
in the form
\begin{eqnarray}
\label{eq: 4.1}
\frac{\partial P}{\partial t}(x,v,t)&=&-v\frac{\partial P}{\partial x}(x,v,t)-\Gamma(v)P(x,v,t)
\nonumber \\
&+&\int_{-\infty}^\infty \mathrm{d}v'\ 
R(v,v')P(x,v',t)
\end{eqnarray}
where $R(v,v')$ is the rate for scattering from velocity $v'$ to $v$, and 
$\Gamma(v)=\int_{-\infty}^\infty \mathrm{d}v'\ R(v',v)$. 
Eq.~(\ref{eq: 4.1}) manifestly reduces to Eq.~(\ref{eq:evol_P}) when only 
two velocities are possible. Therefore,
the analysis for Eq.~(\ref{eq: 4.1}) follows the same steps as for 
Eq.~(\ref{eq:evol_P}), except that 
operations involving matrix multiplication are replaced by integral transforms. 
The key stages in the argument are unchanged, and we conclude that 
in the general case the Boltzmann equation will predict that $\alpha_{\rm E}\ne \alpha$.

In sedimentation problems, however, we are usually 
concerned with the equilibrium of colloidal particles, which are much larger than the 
size of the atoms. Because the mass ratio is very large, the changes in the velocity 
of the colloidal particle with each collision are small.
This can be described by replacing 
the general collision term in the Boltzmann equation (\ref{eq: 4.1}) with a 
diffusion term. Specifically the velocity of the particle undergoes diffusive fluctuations,
with diffusion coefficient ${\cal D}$, while relaxing to a drift velocity $v_0$ with rate constant
$\gamma$, so that $v$ obeys the stochastic differential equation
\begin{equation}
\label{eq: 4.2}
\mathrm{d}v=-\gamma(v-v_0){\rm d}t+\sqrt{2\mathcal{D}}\mathrm{d}\eta
\end{equation}
where $\langle \mathrm{d}\eta\rangle =0$ and $\langle \mathrm{d}\eta^2\rangle=\mathrm{d}t$.
The corresponding Fokker-Planck equation is
\begin{equation}
\label{eq: 4.3}
\frac{\partial P}{\partial t}=-v\frac{\partial P}{\partial x}+\gamma\frac{\partial}{\partial v}\left[(v-v_0)P\right]
+{\cal D}\frac{\partial^2 P}{\partial v^2}
\end{equation}
where the collision kernel in (\ref{eq: 4.1}) has been replaced by a diffusion term.
This is a variant of the Ornstein-Uhlenbeck process \cite{Uhl+30}, which is 
often an accurate description of the velocity of a Brownian particle (one then 
speaks of a Rayleigh particle \cite{Hoa+71}).
The normalisable steady-state solution of the Fokker-Planck equation (\ref{eq: 4.3}) is
\begin{equation}
\label{eq: 4.4}
	P(x,v) \propto \exp\left(-\frac{\gamma}{2\mathcal{D}}v^2\right) \exp\left(\frac{v_0\gamma^2}{\mathcal{D}}y\right)
\ .
\end{equation}
Determining the spatial diffusion coefficient for the process described 
by (\ref{eq: 4.2}) gives $D={\cal D}/\gamma^2$, so that (\ref{eq: 4.4}) agrees with (\ref{eq: 5}).
In fact, as the process described by (\ref{eq: 4.3}) is Gaussian, the expansion in (\ref{eq:cum_exp})
reduces to its two first terms, and thus one has $\alpha=\alpha_{\rm E}$. 
We conclude that the classical Einstein relation for sedimentation equilibrium 
is valid for macroscopic colloidal particles, while it may fail for microscopic 
particles with a mass which is comparable to that of the gas.

\section{ Conclusions }
\label{sec: 5}

In this letter, we have investigated a class of stochastic problems, with
a mean drift, and a reflecting wall. This corresponding to the classical
and fundamental problem of sedimentation equilibrium \cite{Perrin:1913,Einstein:05,Sutherland:05}.
Very general considerations lead to the
conclusion that the distribution, far away from the wall, decays exponentially.
We have shown that the decay rate, $\alpha$, can be determined quite simply 
from large deviation theory using equation (\ref{eq: 3.2}) (where the cumulant
generating function is available) or equation (\ref{eq:cum_exp}) (when the cumulants
of $f(t)$ are known). Whereas the classical Einstein relation can
be derived from a Fokker-Planck description of the evolution of the PDF,
our approach does {\it not} rest on any Fokker-Planck description.

In the case of the telegraph noise model we show explicitly that 
$\alpha\ne\alpha_{\rm E}$. This raises the question as to whether 
there is a reason to doubt the validity of the Einstein relation for
sedimentation equilibrium properties. We have argued that 
while $\alpha$ need not equal $\alpha_{\rm E}$ for the 
Boltzmann equation, in the limit where the ratio of the mass of the 
suspended particles is very large, the Boltzmann equation should be 
replaced by a variant of the Ornstein-Uhlenbeck model. An explicit solution 
shows that $\alpha=\alpha_{\rm E}$ for this case.

Lastly, it is of interest to note that in some cases the scaled cumulant generating function $\lambda(k)$
does not exist, for example then the process is discrete in time $x_{n+1}=x_n+f_n$ and the velocities
are independent and follow the density of probability $p(f_n)\propto(1+|f_n-v_0|)^{-\beta}$ with $\beta>3$.
Then both $\langle f\rangle=v_0$ and $D$ are finite, but $\lambda(k)$ is nowhere defined, except in $k=0$.
The large tails in the distribution of $f_n$ avoid to properly define a region in space where the dynamic is
considered being far from the wall, as the particles are likely to do large jumps. It is then related to the 
mean free path interpretation of the telegraph model; in this case not only the Einstein relation, but also 
the exponential sedimentation are no longer valid, and long range corrections must be added.

Acknowledgements: MW thanks the Kavli Institute for Theoretical 
Physics, Santa Barbara, where work on this paper was supported in part by the 
National Science Foundation under Grant No. NSF PHY11-25915.


\end{document}